\documentclass[aps,groupaddress,prb,showpacs]{revtex4}

\usepackage{graphicx}
\usepackage{bm}
\usepackage{graphicx}
\usepackage{amssymb}
\usepackage{bm}
\usepackage{hyperref}

\def\xp{$X^+$ }
\def\xm{$X^-$ }
\def\x0{$X^0$ }

\begin{document}

\title{Hyperfine interaction in InAs/GaAs self-assembled quantum dots :
 dynamical nuclear polarization versus spin relaxation }


\author{O.~Krebs}
\email[Corresponding author : ]{Olivier.Krebs@lpn.cnrs.fr}
\author{B.~Eble}
\author{A.~Lema\^{i}tre}
\author{P.~Voisin }
\affiliation{CNRS-Laboratoire de Photonique et de Nanostructures, Route de Nozay, 91460
Marcoussis, France}
\author{B. Urbaszek}
\author{X. Marie}
\author{T. Amand}
\affiliation{Laboratoire de Physique et Chimie des Nano-Objets, INSA/CNRS/UPS, 135 avenue de Rangueil, 31077 Toulouse Cedex 4, France}

\begin{abstract}
We report on  the influence of hyperfine interaction on the optical orientation of singly charged excitons $X^\pm$ in self-assembled InAs/GaAs quantum dots. All measurements were carried out on individual quantum dots studied by micro-photoluminescence at low temperature. We show that the hyperfine interaction leads to an effective partial spin relaxation, under 50~kHz modulated excitation polarization, which becomes however strongly inhibited under steady optical pumping conditions because of dynamical nuclear polarization. This optically created magnetic-like nuclear field can become very strong (up to $\sim$4~T) when it is generated in the direction opposite to a longitudinally applied field, and exhibits then  a bistability regime. This effect is very well described by a theoretical model derived in a perturbative approach, which reveals the key role played by the energy cost of an electron spin flip in the total magnetic field. Eventually, we emphasize the similarities and differences between \xp and  \xm trions with respect to the hyperfine interaction, which turn out to be in perfect agreement with the theoretical description.\\
\end{abstract}

%\keywords{}

\maketitle

\section{Introduction}
\label{Intro}
Manipulating quantum coherence in condensed matter at the nanometer scale is a very exciting challenge. In this respect the spin degree of freedom of electrons trapped in semiconductor quantum dots (QDs) presents many interesting features. It is weakly coupled to the fluctuating environment of a quantum dot ensuring long coherence and relaxation times~\cite{PRL89-Cortez,Kroutvar-Nature432,Sci313-Greilich}, can be investigated by optical methods~\cite{PRL94-Laurent,PRB68-Flissikowski}, and has recently proven to be coherently controllable  by a r.f. field~\cite{Koppens-Nature442,kroner-2007}. Yet a major issue for such coherent manipulation originates from the hyperfine interaction with the nuclear spins of the host material~\cite{Optical-Orientation,PRB65-Merkulov2002}, which acts as a random magnetic field, and is thus a source of spin dephasing. This is particularly true for III-V's-based quantum dot, since all nuclei have a non-zero spin in these materials. Another related aspect is the dynamical polarization of the nuclei~\cite{PRL86-Gammon-Merku,PRB-Eble,PRL96-Lai,PRL97-Akimov} which occurs when the electron spin polarization is maintained out of equilibrium thanks to an efficient optical pumping. Depending on the external magnetic field, this process may give rise to a large magnetic-like field  which  affects the electron spin dynamics. In this article, we present our recent  investigations of optical orientation of singly charged excitons (trions) in InAs/GaAs quantum dots which have revealed the prominent role played by the electron-nuclei hyperfine coupling.\\

\section{Optical spectroscopy of singly charged excitons}

\begin{figure}[h]\centering
\includegraphics[width=12 cm,angle=0]{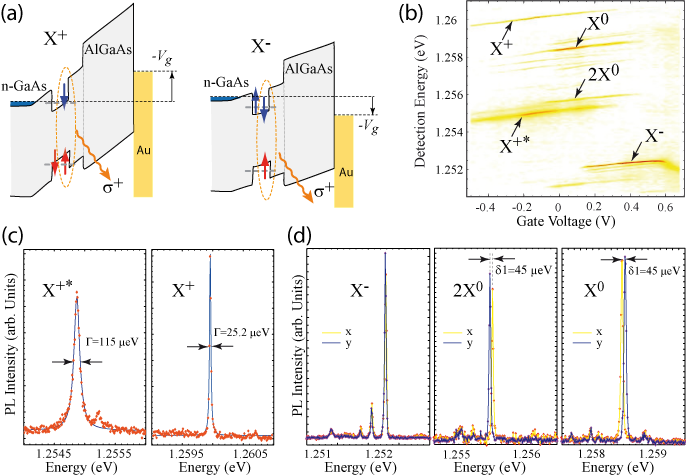}
\caption{ (a) Schematics of the band profile along the growth direction of a n-Schottky structure, illustrating the photoluminescence of positive or negative trions depending on the applied gate voltage $V_g$. (b) Contour plot of the PL intensity from a single InAs QD at T$\approx$10~K versus the detection energy and applied gate voltage. Excitation energy is  1.34~eV. (c) PL spectra at $V_g=-0.24$~V showing the difference of linewidth for   the ground ($X^+$) and hot ( $X^{+^\star}$) positive trions. (d) PL spectra at $V_g=+0.3$~V measured in two ortogonal linear polarizations ($x$ and $y$) showing the  mirrored fine structure splitting $\delta_1$ of the neutral biexciton ($2X^0$) and exciton ($X^0$).
} \label{SchottkyDiode}
\end{figure}

We have studied InAs self-assembled quantum dots grown by molecular beam epitaxy on a (001)-oriented semi-insulating GaAs substrate.  These quantum dots which are formed in the Stranski-Krastanov mode, are lens-shaped with  a typical  20~nm diameter  and a 4~nm height. Two types of samples were fabricated and investigated : (i) charge-tuneable QDs  embedded in  the intrinsic region of a n-i-Schottky diode, (ii) singly charged QDs due to residual p-type doping.  The charge-tunable samples consist of a single QD layer grown 25~nm above a 200~nm-thick n-GaAs layer and covered by  GaAs~(25~nm)$\backslash$Al$_{0.3}$Ga$_{0.7}$As~(120~nm)$\backslash$GaAs~(5~nm). In this case the QD charge is controlled by an electrical bias  applied between a top  semi-transparent Schottky contact  and a back ohmic contact. This is schematically illustrated in Fig.~\ref{SchottkyDiode}(a). We used  metallic shadow masks evaporated on the sample surface with 1~$\mu$m-diameter optical apertures to   select spatially single QDs. This technique was however not  systematically employed because in most  cases the QD density was low enough to enable us the study of individual spectral lines in the low energy tail of the QD distribution.\\
\indent The $\mu$-photoluminescence spectroscopy of individual InAs  QDs was carried out  in a standard confocal geometry with an optical excitation provided by a cw  Ti:Sapphire laser. Most of the experiments reported here were performed with a  magneto-optic cryostat working in the Faraday geometry, namely with the magnetic field applied parallel to the optical axis. In this case, a 2~mm focal length aspheric lens (N.A.~0.5) was used to focus the  excitation beam and to collect the PL from the sample, while the relative positioning of the sample in all three directions  was achieved by  piezo-motors. This very compact microscope  was installed  in the cryostat insert ensuring thus an excellent  mechanical stability as requested for the study of individual quantum dots. The PL beam was dispersed by a  double spectrometer of 0.6~m-focal length and detected by a Nitrogen-cooled CCD array camera.\\
The optical selection rules of the ground interband transition are determined by the \emph{heavy} character of the confined hole.  In these flat and biaxially strained quantum dots, the hole ground state can  be considered as a pure heavy-hole well separated from the light-hole states. It is  described  by the projection of its  angular momentum $m_z=\pm3/2$ along the growth axis $z$, while its envelope wave-function retains a $S$-like character. As a result it gives rise to pure selection rules for the optical recombination of an electron-hole pair (named further exciton), namely the $|\pm1\rangle\equiv|s_z=\mp1/2,m_z=\pm3/2\rangle$ states emit $\sigma\pm$~circularly polarized photons while the  $|\pm2\rangle\equiv|s_z=\pm1/2,m_z=\pm3/2\rangle$  states are not coupled to light. As usually done, we thus define the circular polarization of the photoluminescence (PL) signal collected along the $z$ axis  by :
\begin{equation}
\mathcal{P}_c=\frac{I_{PL}^{\sigma+} - I_{PL}^{\sigma-}}{I_{PL}^{\sigma+} + I_{PL}^{\sigma-}}
\end{equation}
where $I_{PL}^{\sigma\pm}$ is the PL intensity measured in $\sigma\pm$ polarization. Such measurement enables us to directly monitor either the electron spin polarization for positively charged trions $X^+$, or the hole spin polarization for negatively charged trions $X^-$ as shown in  Fig.~\ref{SchottkyDiode}(a). For  neutral excitons (denoted $X^0$), the $|\pm1\rangle$ bright states are generally split by the electron-hole exchange into linearly-polarized states because of the lack of perfect rotational symmetry. The corresponding splitting $\delta_1$ amounts to a few tens of $\mu$eV, so that the circular polarization averages to zero over the radiative lifetime $\tau_r\gg\hbar/\delta_1$ because of the quantum beats between the $|+1\rangle$  and $|+1\rangle$ states. Noteworthily, this effect does not take place for the trions $X^\pm$ because in this case an unpaired  carrier (electron or hole) interacts with a pair of carriers (holes or electrons) in a singlet configuration, which cancels out the  exchange interaction.  The spin degeneracy  of the ground trion states is  actually a  consequence of the Kramer's degeneracy expected for half-integer spin system in zero magnetic field. The charge control of a quantum  dot, or at least its determination is therefore a crucial aspect of optical orientation in self-assembled quantum dots.\\
\indent Figure~\ref{SchottkyDiode}(b) shows an example of such a charge tuning in a single quantum dot. The neutral exciton ($X^0$) and biexciton  ($2X^0$) lines are determined from their fine structure shown in Fig.~\ref{SchottkyDiode}(d). The negative trion $X^-$ is formed when the built-in electric field is reduced so that the dot gets filled with one or two resident electrons provided by the  n-GaAs layer \cite{PRL94-Laurent,PRL79-Warburton}. Less intuitive, positive trions can be also generated  in such sample thanks to the optical charging of the QD valence state with a resident hole. This complex is  formed  by decreasing  the gate voltage, namely increasing the internal electric field,  until the neutral exciton line ($X^0$) vanishes (see Fig.~\ref{SchottkyDiode}(b))   and by using ``intra-dot'' excitation conditions to photo-create holes in the QD~\cite{APL86-Ediger}. Here a spectral line $X^{+^\star}$  identified as a hot trion (namely a trion \xp with one hole on a $P_h$ orbital) is also visible. As emphasized in Fig.~\ref{SchottkyDiode}(c), it is featured  by a Lorentzian lineshape of 115~$\mu$eV width much larger than the $\sim$25~$\mu$eV linewidth measured   for $X^{+}$ (limited here by the spectral resolution), as a consequence of the short lifetime ($\sim$10~ps) of the final state, namely a single hole in an excited level.

\section{Spin relaxation induced by the hyperfine coupling}
\label{e-N_coupling}
 \indent Due to the $P$ symmetry of the Bloch wavefunctions in the valence band, the coupling of the nuclei with holes can be generally neglected because the Fermi contact interaction vanishes.  This indeed leads  to  very distinctive spin dynamics between \xp and \xm trions, as it will be shown below. The Hamiltonian describing the hyperfine interaction of a  conduction band  electron  (spin $\hat{\bm{S}}^e=\frac{1}{2}\hat{\bm{\sigma}}$)  with the $N$ nuclear spins  of a quantum dot is given by \cite{Optical-Orientation,PRL86-Gammon-Merku}:
 \begin{equation}\label{H-hyperfin} \hat{H}_{hf}=\nu_0\sum_{j}A_j\left|\psi(\mathbf{r}_j)\right|^2 \hat{\bm{I}}_j\cdot\hat{\bm{S}}^e
 \end{equation}
where $\nu_{0}$ is the two-atom unit cell volume, $\mathbf{r}_{j}$ is the position of the nuclei $j$ with spin $I_{j}$, $A_{j}$ is the constant of hyperfine interaction with the electron and $\psi(\mathbf{r})$ is the electron envelope wave-function. The sum runs over the $N$ nuclei interacting significantly with the electron (i.e. in the effective QD volume defined by $V= (\int \left|\psi(\mathbf{r})\right|^{4}d\mathbf{r})^{-1}=\nu_{0}N/2$). To zeroth order, the role of the hyperfine interaction is equivalent to an effective  magnetic  field  $\bm{B}_n$ acting  on the electron spin :
 \begin{equation}\label{NucField} \bm{B}_n=\frac{\nu_0\sum_{j}A_j\left|\psi(\mathbf{r}_j)\right|^2 \langle \bm{I}_j\rangle}{g_e \mu_B}
 \end{equation}
 where $\mu_B$ is the Bohr magneton and $g_e$ is the effective electron Land\'{e} factor in the QD. Under certain conditions that will be discussed further, this magnetic field can reach a considerable strength up to a few Tesla in   InAs QDs, which obviously affects in turn the spin dynamics of a confined electron.  Besides, even when conditions are met to keep its average value to zero, it still presents statistical fluctuations  due to the finite number $N\sim10^5$ of involved nuclei, which typically amounts to a few tens of  milli-Tesla. This yields electron spin precession in a characteristic time $T_\Delta\sim$500~ps which is responsible for spin dephasing   in an  ensemble of charged quantum dots. This effect was evidenced in Ref.~\cite{PRL94-Braun} for an ensemble of p-doped QDs for which the photoluminescence of positive trions \xp (one electron with two holes) directly monitors the  average electron spin component $\langle \hat{S}_z^e \rangle$.\\
\indent For  \xp trions,  the nuclear field $\bm{B}_{n}$ can be considered as \textit{frozen} because its correlation time $T_{2}$$\approx$10$^{-4}$s (determined by the dipole-dipole interaction between nuclei) is much longer than the excitonic radiative lifetime of the trion $\tau_{r}\approx~0.8$~ns. When the nuclear field fluctuation $\delta\bm{B}_n(t)=\langle\bm{B}_n(t)\rangle_{\tau_r}$ is parallel to the optical axis $z$ along which the spin is optically written no relaxation occurs. The apparent spin relaxation of \xp  results from the random orientation of   $\delta\bm{B}_n(t)$  over the integration time of the measurement. In the experimental studies reported here the condition for assuming such a random nuclear field orientation is fulfilled  since the integration time ($\sim$1\--10~s) is  few orders of magnitude longer than $T_{2}$. As a result, the time-integrated circular polarization of a single \xp line excited with circularly polarized light is given by:
\begin{equation}
\mathcal{P}_c=2\int\langle \hat{S}_z^e(t)\rangle\exp(-t/\tau_{r})/\tau_{r}dt\label{Eq:polar}
\end{equation}
 where $\langle \hat{S}_z^e(t)\rangle$ is the electron spin evolution averaged over the distribution of random nuclear fields. Using the expression derived in Ref.~\cite{PRB65-Merkulov2002} for $\langle \hat{S}_z^e(t)\rangle$ with a characteristic \textit{r.m.s.} nuclear field $\Delta_{B}\approx 30$~mT, we find that for an initially photocreated spin $S_z^e(0)=1/2$ the maximum degree of polarization which can be reached amounts to $\mathcal{P}_c^{max}$=53\%. It is noteworthy that the average spin evolution  described by $\langle \hat{S}_z^e(t)\rangle$ in Ref.~\cite{PRB65-Merkulov2002} does not decay to zero, but to 1/3 of its initial value in the time of a few $T_\Delta$. This is due to the  non-zero nuclear field fluctuation component along $z$ leading to the average projection factor  $\int\cos^2\theta=1/3$. Here, taking into account the \xp finite lifetime $\tau_r$ simply  reduces the  effective spin relaxation from 2/3 to about one-half.\\
\begin{figure}[h]\centering
\includegraphics[width=12 cm,angle=0]{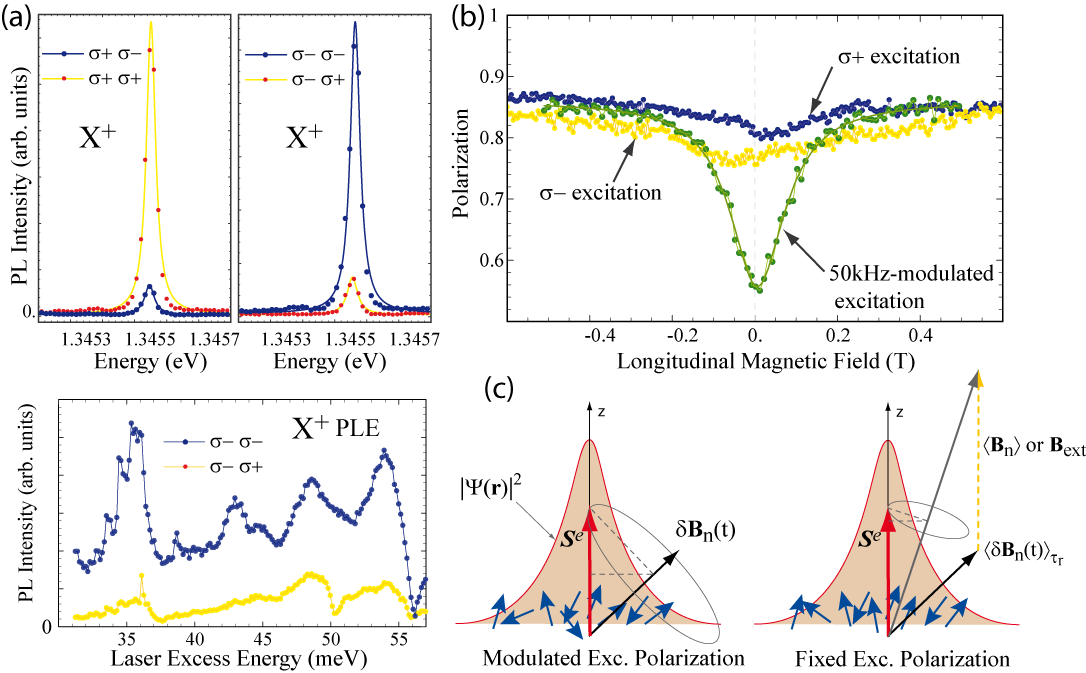}
\caption{(a) PL spectra of a single \xp line measured in zero magnetic field at $T=5$~K under circularly polarized excitation and detection as indicated. The laser energy was set to an excitation resonance at 1.38 eV likely associated to 1LO-phonon assisted transition as shown in the PLE spectrum. Solid lines are the Lorentzian fits of the experimental data. (b) Circular polarization of the \xp PL intensity as a function of a longitudinal magnetic field. Excitation polarization is either fixed ($\sigma^+$ and $\sigma^-$) or modulated at 50~kHz between $\sigma^+$ and $\sigma^-$ by a photo-elastic modulator. (c) Schematics of the spin relaxation induced by the nuclear field fluctuations $\delta\bm{B}_n(t)$ depending on the mode of excitation (fixed or modulated polarization).} \label{FigSpinRelax}
\end{figure}
 \indent Figure~\ref{FigSpinRelax} presents typical  results of optical orientation for  an \xp complex under quasi-resonant excitation, namely with a laser energy at +35~meV above the \xp emission line. This resonance in the excitation spectrum can be reasonably ascribed to a LO-phonon assisted excitation of the trion in its ground state, which offers almost ideal conditions for spin conservation.  The first observation is that the degree of circular polarization that can be achieved under circularly polarized excitation is actually much higher than the above-mentioned theoretical maximum, with $\mathcal{P}_c\sim80\,\%$ in zero magnetic field.  This surprising result is elucidated by replacing the fixed laser polarization by  a 50~kHz-modulated polarization between $\sigma^+$ and $\sigma^-$. The latter is provided by a photo-elastic modulator, while the signal of $\sigma^+$ polarized PL is measured  by a Si-avalanche photodiode with time-gating synchronized to the modulation. In zero magnetic field, we observed a dramatic drop of the PL circular polarization for $X^+$, from $\sim$85\% (under steady $\sigma^+$  excitation polarization) down to $\sim$55\%. This striking result points toward the building of a  nuclear polarization  under constant excitation polarization, giving rise to a non-zero average nuclear field $\bm{B}_{n}$ parallel to the optical axis. The direction of electron spin precession of \xp acquires thus a finite component along $z$ so that  the spin relaxation vanishes for $|B_{n}|\gg\Delta_B$ as illustrated in Fig.~\ref{FigSpinRelax}(d). Under $\sigma^{+}/\sigma^{-}$~alternative excitation,  such a nuclear polarization cannot settle because its characteristic risetime $T_{1e}$ is much longer than the modulation period of 20~$\mu$s~\cite{Optical-Orientation,Maletinsky-PRL99}. To support this interpretation we have  measured the dependence of $\mathcal{P}_c$ on a longitudinal magnetic field $B_{z}$ allowing the progressive suppression of hyperfine-induced spin relaxation (HSR)~\cite{PRB65-Merkulov2002,PRL94-Braun}. Fig.\ref{FigSpinRelax}(b) shows that for $|B_{z}|\geq $200~mT the circular polarization of \xp is restored at $\sim$85\%, i.e. to the same level as under steady polarization  excitation. Qualitatively, this confirms that the drop of $\mathcal{P}_c$ in zero-magnetic field is really due to an effective random magnetic field of the order of a few tens of  milli-Tesla. Yet, the half-width of the polarization dip amounts to $\sim$80 mT which is about 3 times larger than the expected nuclear field fluctuations $\Delta_B$, whereas the reduction of $\mathcal{P}_c$ at $B_z=0$~T is actually only 35\% instead of the expected 50\%. These  observations seem rather  contradictory because they would respectively indicate  a higher or  smaller value for $\Delta_B$. A complete description of HSR is certainly still required, including e.g. the possible spin cooling due to the electron Knight field~\cite{Optical-Orientation}, the quadrupolar interactions between the nuclei~\cite{Korenev-PRL}, as well as a possible small  disequilibrium between the effective $\sigma^+$ and $\sigma^-$ excitation intensities. All in all, the most striking result is that HSR is very efficiently suppressed by the nuclear field that it gives rise to under fixed excitation polarization.\\
\begin{figure}[h]\centering
\includegraphics[width=12 cm,angle=0]{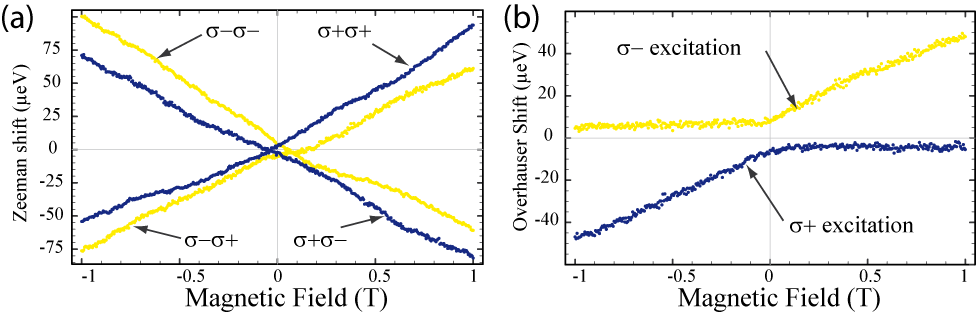}
\caption{(a) Zeeman shift of both $\sigma^\pm$  components of the \xp line for both $\sigma^\pm$ excitation polarizations. (b) Overhauser shift of \xp with respect to a linear Zeeman splitting measured  for  both $\sigma^\pm$ excitation polarizations.} \label{FigOHS1}
\end{figure}
%In the case of  $X^-$, the influence of the nuclear field is less important  because once the trion has relaxed on its ground state its polarization is determined by   the hole spin which has a negligible hyperfine interaction with the nuclei. This explains why no significant change of polarization is observed between both modes of excitation.\\

 In this regime,  the building of a significant nuclear polarization requires in principle  a small external magnetic field to inhibit the nuclear depolarization due to the dipolar interaction between nuclei in the time $T_{2}\sim100\,\mu$s~\cite{Optical-Orientation}. In InAs QDs, this requirement seems to be raised  because of the strong effective magnetic field produced by a spin polarized electron (or Knight field) on the  QD nuclei.  Due to the weight of the envelope function in Eq.~\ref{H-hyperfin}, this field is actually  non-uniform, but its  typical strength  given by $B_e\sim 2\langle S^e_z\rangle \tilde{A}/\hbar\tilde{\gamma}_n N$  amounts to ~$\sim$10~mT.  This  turns out to be  quite sufficient to inhibit the nuclear depolarization in zero external field~\cite{PRL96-Lai,Maletinsky-PRL99,Oulton-PRL98}, and probably explains the weak dips of $\mathcal{P}_c$ visible in Fig.~\ref{FigSpinRelax}(b) for the measurements under fixed polarization. These dips which are indeed shifted by a positive or negative magnetic field depending on the sign of excitation polarization ($\sigma^+$ or $\sigma^-$) could result from a reduction of the nuclear field when $B_z=-B_e$,  yielding thus a more efficient HSR.\\
\indent To demonstrate that a significant nuclear field $B_{n}$ is  optically induced in the QD, the electron spin splitting  due to the nuclear field  $\delta_{n}=g_e\mu_{B}B_{n}$, also called Overhauser shift,  can be straightforwardly measured by polarization resolved PL spectroscopy~\cite{PRB-Eble,PRL86-Gammon-Merku,PRL96-Lai,PRB71-Yokoi,Braun-PRB74,Maletinsky-PRB75,Tartakovskii-PRL98}. Such measurements are reported in Fig.~\ref{FigOHS1} for the same \xp trion as in Fig.~\ref{FigSpinRelax}. The splitting between the $\sigma^+$ and $\sigma^-$ PL components is  then given by $\delta_Z-\delta_n$ where $\delta_Z=(g_h-g_e) \mu_B B_z$ is the Zeeman splitting for the electron-hole pair involved in the \xp transition. The finite nuclear field is revealed by the shift of \xp zero splitting  to a finite magnetic field which is either positive or negative depending on the excitation polarization. The Overhauser shift  $\delta_n$ that can be  deduced from both these measurements is reported in Fig.~\ref{FigOHS1}(b).   The strength of nuclear field $|B_n|=|\delta_n/g_e\mu_B|$ amounts to $\sim$200~mT  in zero external field for  a characteristic  electron Land\'{e} factor $|g_e|=$0.5. This agrees with the strong inhibition of HSR in steady excitation polarization regime. More spectacular is  the  dependence of $\delta_n$ with the applied magnetic field revealed by this plot. Under a given excitation polarization, the optically generated nuclear field either increases linearly with the applied field in  one direction, or  remains almost constant in the opposite direction. This  behavior points toward a non-linear  mechanism for the optical generation of nuclear polarization. As we will see in the next section, the key parameter to explain this asymmetry and this non-linear dependence is the total electron spin splitting $|g_e(B_z+B_n)|$ which represents the energy cost of each electron-nucleus flip-flop, and which obviously depends on the respective sign of the external and nuclear fields.\\

\section{Dynamical Nuclear Polarization}

\subsection{Theoretical model}

\indent The optically induced spin polarization of the nuclei results from successive electron-nucleus ``flip-flops'' mediated by the hyperfine interaction. To derive from Eq.~(\ref{H-hyperfin}) a tractable expression for the nuclear polarization dynamics we first assume a uniform electron wavefunction $\psi(\mathbf{r})=\sqrt{2/N\nu_{0}}$ spanning over  $N$  nuclei in the quantum dot. The  flip-flop  term  included in the Hamiltonian~(\ref{H-hyperfin}) reads then   $\hat{H}_{ff}=\frac{\tilde{A}}{2N}(\hat{I}_+\hat{S}^e_-+ \hat{I}_-\hat{S}^e_+)$. By treating this coupling  as a random  time-dependent perturbation (with zero mean), one can derive straightforwardly  the    expression for the electron-induced relaxation time $T_{1e}$  of a given nuclear spin~\cite{Abragam,Optical-Orientation,PRB-Eble}:
\begin{equation}
T_{1e}=\left(\frac{N\hbar}{\tilde{A}}\right)^2\frac{1+(\Omega_e\tau_c)^2}{2f_e\tau_c}\label{Eq:T1e}
\end{equation}
where $\hbar\Omega_e=g_e\mu_B(B_z+B_n)$ is the electron spin splitting, $\tau_c$ is the correlation time of the perturbation $\hat{H}_{ff}$, and $f_e$ is the  fraction of time that the QD contains an unpaired electron. In the above expression, we also introduced a nucleus-independent hyperfine interaction constant $\tilde{A}$, since the $A_j$'s  vary weakly among the different isotopic species of InAs QDs. The expression obtained for $T_{1e}$ is thus independent on the nuclear species which enables us to derive in the approximation of high nuclear spin temperature~\cite{Abragam} a simple rate equation for the average nuclear spin $\langle I_z \rangle=\sum_j\langle \hat{I}_{j,z}\rangle$ in a quantum dot :
\begin{equation}\label{eq:DNP}
\frac{d\langle I_z\rangle}{dt}=-\frac{1}{T_{1e}}\left[\langle I_z\rangle-\widetilde{Q}\left(\langle \hat{S}_z^e\rangle-\langle \hat{S}_z^e\rangle_{_{  0}}\right)\right] - \frac{\langle I_z \rangle}{T_d}
\end{equation}
where  $\widetilde{Q}=\sum_jI_j(I_j+1)/(NS(S+1))$ is a numerical factor estimated to $\sim15$ in actual In$_{1-x}$Ga$_x$As QDs containing a fraction $x\sim0.5$ of Gallium, $\langle \hat{S}_z^e\rangle_{_{  0}}$ is the average electron spin at thermal equilibrium,  and $T_d$ is an effective time constant introduced here to described the losses of nuclear polarization. Such term is quite necessary because, in its absence, the stationary solution of Eq.~(\ref{eq:DNP}) driven by the average electron spin $\langle \hat{S}_z^e\rangle$  would lead to a much higher nuclear polarization than observed in our experiments with trions. Different mechanisms likely contribute to this effect, e.g.  the dipolar interaction between nuclei responsible for fast  depolarization in a very weak field and for a slower field-independent spin diffusion, or the  quadrupolar coupling with local electric field gradients which could be the prominent term in InAs QDs because of the local anisotropic strains~\cite{PRB-Paget-Amand}. Besides,  the temporal fluctuations of the Knight field $\propto(\hat{S}_z^e-\langle \hat{S}_z^e\rangle)$ are susceptible to enhance these depolarization mechanisms~\cite{PRL86-Gammon-Merku}.\\
\indent Equations (\ref{Eq:T1e}) and (\ref{eq:DNP})  clearly show  the  feedback of the nuclear field on its effective building rate $(T_d+T_{1e})/T_d T_{1e}$ via the electron spin splitting $\hbar \Omega_e$. The latter originates from the difference in energy of the electron-nucleus levels involved in  flip-flop transitions, which is indeed principally determined by the Zeeman electron splitting in the total field $B_n+B_z$. The related issue of energy conservation in spin flip-flop processes is  sorted out  by the level broadening  $\hbar/\tau_c$ associated to the finite correlation time of the hyperfine coupling~\cite{Urbaszek-PRB76r}, as illustrated in the inset of Fig.~\ref{FigOHS2}.  The dependence of $T_{1e}$ on $\hbar\Omega_e$ is directly responsible for the strong asymmetry on the magnetic field dependence (see Fig.~\ref{FigOHS1}) : if the external field $B_z$ is applied parallel to the nuclear field, the splitting $\hbar \Omega_e$ increases accordingly, limiting considerably the strength of $B_n$, even though the depolarization rate $T_d^{-1}$ is reduced  in the same time. On the opposite, if the magnetic field is applied antiparallel to $B_n$ the electron splitting is reduced and  the nuclear polarization rate increases. This gives rise  to a positive feedback on $B_n$ such that $|B_n|>|B_z|$. This relation holds  until a critical field  for which the polarization losses $-\langle I_z\rangle/T_d$ which are proportional to $B_n$  overcome the maximum polarization rate (at $\Omega_e=0$). Eventually, the feedback due to the electron spin splitting is the cause of strong non-linearities as a function of external parameters (magnetic field, electron spin polarization, excitation power...)   giving rise generally to spectacular  bistability regimes~\cite{Braun-PRB74,Tartakovskii-PRL98,Maletinsky-PRB75,Urbaszek-PRB76r,Kaji-APL91}.

\subsection{Non-linearity and bistability in  magnetic field}

\begin{figure}[h!]\centering
\includegraphics[width=12 cm,angle=0]{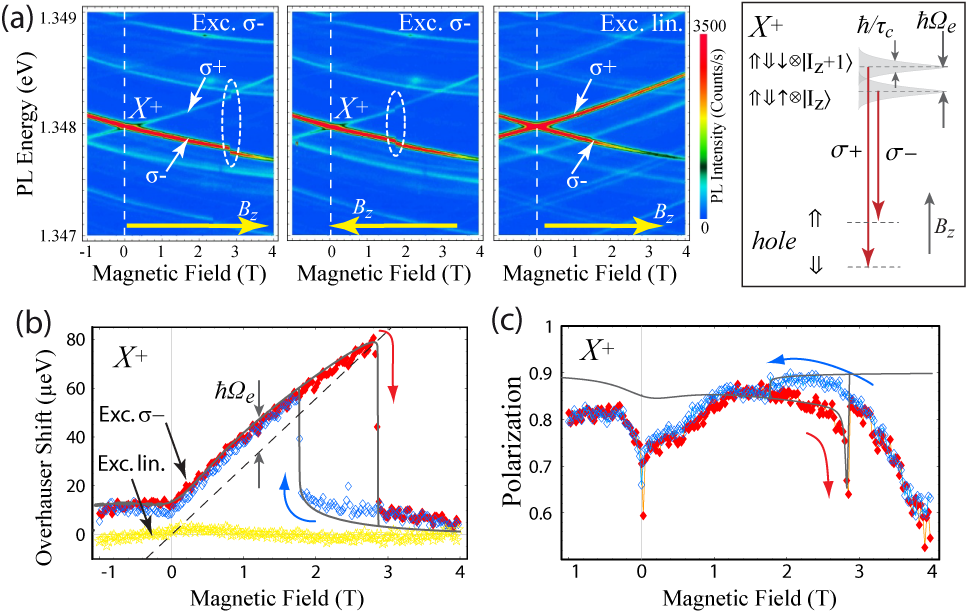}
\caption{(a) PL intensity contour plot of  a \xp positive trion as a function of longitudinal magnetic field and energy detection at $T=2$~K. The excitation is either $\sigma^-$   or linearly  polarized, and the sweep direction  of the magnetic field $B_z$ is indicated by  horizontal arrows for each plot.  (b) Overhauser shift of the \xp line determined from the trion  $\sigma^+/\sigma^-$ splitting in (a) after subtracting the theoretical  Zeeman splitting assumed to be perfectly linear in magnetic field. The dashed line corresponds to the electron Zeeman splitting  $|g_e|\mu_B B_z$. (c) Circular polarization of the trion under $\sigma^-$ excitation for both directions of the magnetic field sweep. In (b) and (c) the solid lines correspond to a fit by the model of dynamical nuclear polarization.}  \label{FigOHS2}
\end{figure}

\indent Figure \ref{FigOHS2} presents Overhauser shift measurements of a positively charged quantum dot similar to those discussed in Sect.~\ref{e-N_coupling}. The three top panels show the \xp PL spectra plotted on a color scale from -1~T up to 4~T under $\sigma^-$ or linearly polarized excitation. As indicated by  arrows the magnetic field was  either increasing or decreasing. For each field value the PL spectrum was recorded with $\sigma^+$ and $\sigma^-$ analyzer which enables us to measure the effective Zeeman splitting $\delta_Z-\delta_n$ with high precision even in the weak field region. Under   $\sigma^-$ excitation the $\sigma^-$ PL branch is much more intense than the upper  $\sigma^+$~branch, which indicates a successful circular polarization of \xp  achieved thanks to a quasi-resonant excitation  (at 1.38~eV). More remarkable, the \xp splitting  exhibits abrupt changes at 2.8~T and 1.8~T for an increasing or decreasing field respectively,  which clearly indicates a bistability of the nuclear field. On the opposite, the Zeeman splitting  under linearly polarized excitation appears quite linear in $B_z$, proving that no significative nuclear field is created in this case. This result is not that obvious because with $(\langle \hat{S}_z^e\rangle=0$ we still maintain a  discrepancy to the  electron spin  thermal equilibrium  $\langle \hat{S}_z^e\rangle_{_{  0}}$ which can become important in a strong magnetic field. This linear Zeeman splitting  $\delta_Z$ was  subtracted to the  trion splittings measured in all three cases in order to extract  the Overhauser shift $\delta_n$. The results are  reported in Fig.~\ref{FigOHS2}(b) with a theoretical fit provided  by  the model introduced above. To reproduce the behavior at weak field we had to include an explicit field dependence of the depolarization time $T_d$ corresponding to the reduction of nuclear spin relaxation by the nuclear Zeeman effect. Since it is reasonable to assume a quadratic dependence,  we took the following phenomenological expression:
\begin{equation}
\label{Eq:Td}
\frac{1}{T_d}=\frac{1}{T_d^\infty}+\frac{1}{T_d^0}\frac{1}{1+(B_z/B_0)^2}
\end{equation}
The theoretical fit in Fig.~\ref{FigOHS2}(b) was realized by solving numerically the differential equation~(\ref{eq:DNP}) for a magnetic field $B_z(t)$ varying at the experimental sweep rate  of 5~mT/s. This essentially amounts to solve Eq.~(\ref{eq:DNP}) in the stationary regime, with yet  an inherent processing of the hysteresis. We assumed the following QD parameters  $N=5.\,10^4$, $\tilde{A}=50~\,\mu$eV, $Q=13$. The electron $g$ factor  $g_e=-0.48$ could be determined from the collapse point of nuclear field at 2.8~T which occurs when $\hbar\Omega_e=0$, namely when $\delta_n+g_e\mu_B B_z=0$. The optically pumped electron spin $\langle \hat{S}_z^e\rangle=0.45$ was estimated from the measured circular polarization $\mathcal{P}_c\approx0.9$. Eventually, to fit the bistability regime only two parameters had to be varied, the correlation time $\tau_c$ and the product $f_e T_d^\infty$, while the other parameters defining $T_d$  in Eq.~(\ref{Eq:Td}) ($T_d^0$ and $B_0$) could be adjusted afterwards to improve the fit in the weak field region. We found as best fitting parameters, within a tolerance better than 10\%, $\tau_c=42$~ps and $f_eT_d^\infty=1.4$~ms,   $f_eT_d^0=230\,\mu$s and $B_0=0.4$~T. Note that the thermal spin polarization $\langle \hat{S}_z^e\rangle_{_{  0}}\approx 1/2 \tanh(g_e\mu_B B_z/2k_B T)$ plays actually a negligible role in the model because it remains much smaller than the optically created one. In addition, if we take $\langle \hat{S}_z^e\rangle=0$ to treat the case of linearly polarized excitation, we indeed find that the  Overhauser shift  which  develops in a magnetic field because of  $\langle \hat{S}_z^e\rangle_{_{  0}}$  does not exceed 2~$\mu$eV.\\
\indent Since the dynamical nuclear polarization (DNP) consists in a transfer of spin from the \xp unpaired electron to the nuclei, there should be a possible trace of this process on the measured \xp polarization. The flip-flop  rate for the photo-created \xp electron interacting with $N$ nuclei can be deduced from Eq.~(\ref{eq:DNP}) as  $T_{1n}^{-1} = (N\widetilde{Q}/f_e) T_{1e}^{-1}$. When it becomes of the same order of magnitude as the radiative decay rate $\tau_r^{-1}$, the PL polarization should be reduced appreciably. Such an effect is shown in Fig.~\ref{FigOHS2}(c) where a net difference of polarization occurs in the domain of nuclear field bistability, between both branches associated to the increasing or decreasing magnetic field. In particular, when $\hbar\Omega_e$ gets closer to zero a pronounced reduction of $\mathcal{P}_c$  (up to $\approx20$\%) develops until the maximum of nuclear field at 2.8~T, then disappears as soon as the latter vanishes. Yet, the minimum estimate of  $T_{1n}$ with the above fitting parameters gives $\sim$7~ns  which is actually  significantly larger than $\tau_r\sim$1~ns and therefore fails to explain quantitatively this strong reduction. To reproduce the sharp dip of polarization in Fig.~\ref{FigOHS2}(c), we need to consider more specifically the hyperfine induced spin relaxation in the vicinity of $\hbar\Omega_e=0$ as discussed in Sect.~\ref{e-N_coupling}. This effect seems indeed very similar to the Lorentzian dip of $\mathcal{P}_c$ observed with modulated excitation polarization at zero magnetic field (see Sect.~\ref{e-N_coupling}). Here, the same spin dephasing occurs when the total field $B_z+B_n$ vanishes at $\hbar\Omega_e=0$. By adding to $T_{1n}^{-1}$ this first order spin relaxation mechanism as a decay rate $\propto 1/(1+(\hbar\Omega_e/g_e\mu_B\Delta_B)^2)$ the bistability of $\mathcal{P}_c$ can be well reproduced by the spin relaxation term $1/(1+\tau_r/T_{1n})$. Obviously, this simplified  approach does not aim at describing the complete field dependence. In particular, the further decrease of polarization in higher fields is probably due to the increasing spin polarization of the resident hole due to thermalization, an effect which can be observed also under linearly polarized excitation (see Fig.~\ref{FigOHS2}(a)) and which is not taken into account in the simulation. In the weak field region  there is another smooth dip which can be ascribed again to the feedback of $\hbar\Omega_e$ (qualitatively well reproduced), while the very narrow dip at $B_z\approx 0$~T has no clear origin for the moment. It probably results from very specific experimental conditions (e.g. this could be due to electron-hole exchange in the excitation process). This feature  was actually not observed for other investigated quantum dots in contrast to the bistability of $\mathcal{P}_c$, an effect  which demonstrates that the hyperfine induced spin relaxation can survive up to strong magnetic fields ($B_z\gg\Delta_B)$  because of the dynamical nuclear polarization.\\

\subsection{Negative trions}

\begin{figure}[h!]\centering
\includegraphics[width=12 cm,angle=0]{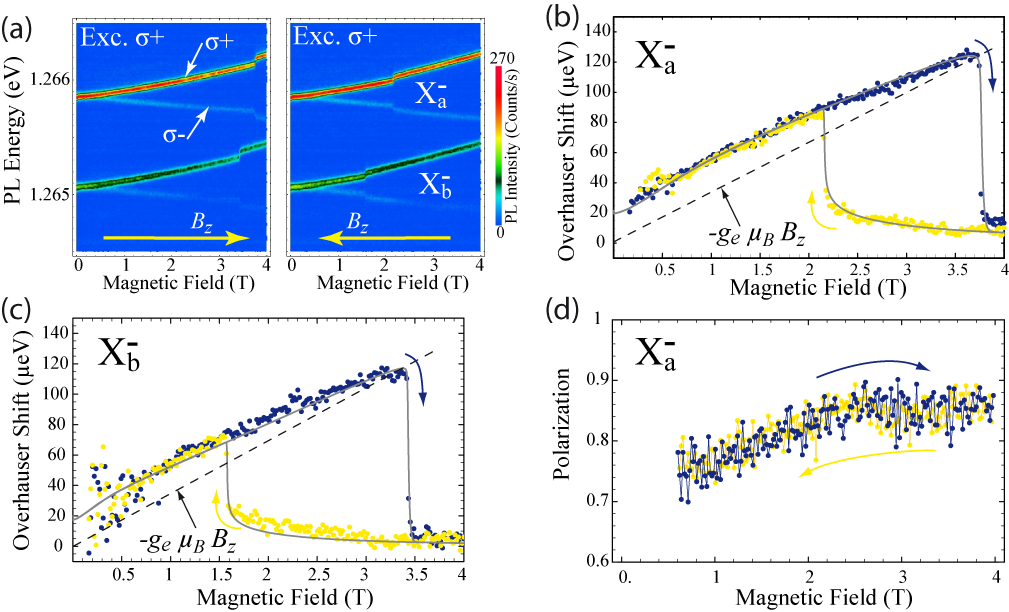}
\caption{(a) PL intensity contour plot of  two   negative trions  denoted $X^-_a$ and $X^-_b$ as a function of longitudinal magnetic field and energy detection at $T=2$~K. The excitation is $\sigma^+$ at an energy of 1.34~eV, and the sweep direction  of the magnetic field $B_z$ is indicated by   horizontal arrows.  (b),(c) Overhauser shift of the \xm lines determined from the  $\sigma^+/\sigma^-$ splitting in (a) after subtracting the theoretical  Zeeman splitting assumed to be perfectly linear in magnetic field. The dashed line corresponds to the electron Zeeman splitting  $-g_e\mu_B B_z$, while the solid line is a fit provided by the model of dynamical nuclear polarization. (d) PL polarization measured for $X^-_a$ as a function  of magnetic field.} \label{FigOHS3}
\end{figure}

A negative trion \xm  consists of an unpaired  confined hole with two conduction  electrons in a singlet configuration (see Fig.~\ref{SchottkyDiode}(a)). Its optical polarization is thus determined by the hole angular momentum ($m_z=\pm3/2$), and therefore should be essentially unsensitive to the nuclear spin system. Under circularly polarized excitation we indeed observe that the PL polarization of \xm remains unchanged when we switch from steady to modulated laser polarization in contrast to \xp trions (see Sect.~\ref{e-N_coupling}). Dynamical nuclear polarization (DNP) is yet quite possible because after emission of a circularly polarized photon  the quantum dot contains a single electron with a well defined spin.  This yields a finite disequilibrium $(\langle \hat{S}_z^e\rangle-\langle \hat{S}_z^e\rangle_{_{  0}})$ during a certain fraction of time $f_e$. Since this remaining electron has a spin orthogonal to the electron recombining with the polarized hole (see Fig.~\ref{SchottkyDiode}(a)) the nuclear field  generated by \xm  should be opposite to that generated by \xp under the same excitation polarization. This effect was clearly evidenced in charge tuneable quantum dots allowing to form either \xm or \xp trions~\cite{PRB-Eble,PRL96-Lai}. Figure~\ref{FigOHS3} reports on the bistability of the nuclear polarization achieved under optical orientation of a negative trion in such a sample at a gate voltage $V_g=+0.5$~V. In the contour-plot of the PL spectra two sets of lines corresponding to two different QDs are visible. The optical excitation is now  $\sigma^+$  to generate a nuclear field antiparallel to the applied field $B_z$. The PL emission exhibits a high  circular polarization $\mathcal{P}_c\approx$80\%. When the magnetic field is increased, the Zeeman splitting shows an abrupt change around 3.8~T and 3.5~T  for $X^-_a$ and $X^-_b$ respectively,  indicating the sudden collapse of  nuclear polarization as in the case of positive trions. When the field is decreased this nuclear field  reappears at 2.2~T ($X^-_a$) and 1.6~T  ($X^-_b$). As in the case of \xp, we obtain a marked bistability regime of the Overhauser shift  which could be fairly well fitted with the same DNP model (Eq.'s~(\ref{Eq:T1e})-(\ref{eq:DNP})). We took the same QD parameters  $N=5.\,10^4$, $\tilde{A}=50~\,\mu$eV, $Q=13$ as for \xp, and $g_e=-0.58\, (-0.6)$, $\langle \hat{S}_z^e\rangle=0.4\,(0.35)$, $\tau_c=27\,(39)$~ps,  $f_e=0.05\,(0.002)$,   $T_d^\infty=0.51\,(0.76)$~s, $T_d^0=0.1\,(0.3)$~s, $B_0= 0.4 \, (0.2)$~T for the $X^-_a$ ($X^-_b$) trion.  Note that the maximum Overhauser shift (125~$\mu$eV) achieved for $X^-_a$  at 3.8~T corresponds to a nuclear polarization of about 50\%. Since the PL spectra were measured only in linear polarization, the Overhauser shift could not be determined with high precision in the weak field domain, which is responsible for the apparent noise in Fig.~\ref{FigOHS3}(b),(c). For the same reason, the PL circular polarization could be determined from Fig.~\ref{FigOHS3}(a) only for magnetic fields above 0.6~T by  fitting separately the intensity of both $\sigma^+$ and $\sigma^-$ branches. The result is reported in Fig.~\ref{FigOHS3}(d) for $X^-_a$. In contrast to positive trions, the nuclear field bistability seems not to affect the PL polarization carried by the hole angular momentum  of \xm, at least in the limit of the experimental noise ($\sim$5\%).  This supports the assumption of negligible hyperfine interaction between hole and nuclei in quantum dots.\\
%\subsection{Non-linearity on electron spin polarization}
\newpage
\section{Conclusion}
In this article, we have presented a few  recent investigations evidencing strikingly  the role played by the hyperfine interaction in the spin physics of an electron confined in InAs/GaAs quantum dots.  Depending on the experimental conditions, this electron-nuclei coupling  gives rise either to significative spin dephasing, or to high nuclear polarization (up to ~50\%), or to a subtle competition between both these effects. These conclusions were drawn from rather straightforward experiments carried out on different individual quantum dots which reveal themselves as a unique system for investigating the hyperfine interaction. Indeed, the  optical selection rules of the interband transitions enable us to obtain very high spin polarization (up to 90\%) under quasi-resonant excitation, and more important, the polarization-resolved micro-photoluminescence spectra provide then a direct means for measuring the average nuclear field, with a precision of $\sim$70~mT, namely about twice its statistical noise. The charge control achieved with  n-Schottky structures  permits to study separately positive or negative trions, and thus to exhibit a remarkable symmetry : depending on the type of trions (\xp or \xm), the spin polarized electron is either in the initial or final state of the trion transition, which determines the sign of the nuclear field created along the optical axis under a given excitation polarization. Another important outcome  is the demonstration of non-linearity and bistability of the dynamical nuclear polarization in quantum dots as a function of an applied magnetic field. The cause of these spectacular effects is linked to the  energy cost of electron-nuclei flip-flops which can be either reduced or enhanced by the magnetic-like nuclear field itself. The regime of strong nuclear polarization is achieved when the external magnetic field is completely compensated by the optically created nuclear field. This internal feedback is very well described by a simple model relying on a uniform hyperfine interaction inside a quantum dot. The remaining issues, which should be addressed now, mostly regards the depolarization mechanisms of nuclear spins, specifically in weak magnetic fields. This seems to us quite necessary to draw up a comprehensive   description of an electron spin in InAs quantum dots.\\
\textbf{Acknowledgements.} This work was partly supported by the ANR contracts BOITQUANT and MOMES.\\

\bibliographystyle{aps}
%\bibliography{BiBlio-Nuclear_Effects}

\end{document}